\documentclass[12pt]{article}
\title{Hidden Source of High-Energy Neutrinos \\
       in Collapsing Galactic Nucleus}
\author{V.~S. Berezinsky$^{a,b}$, V.~I. Dokuchaev$^{b}$ \\
$^{a}${\small\emph
Laboratori Nazionali del Gran Sasso, INFN, Italy} \\
$^{b}${\small\emph
Institute for Nuclear Research, Russian Academy of Sciences, Moscow}}

\begin{document}

\date{}
\maketitle

\begin{abstract}

We propose the model of a short-lived very powerful source of high
energy neutrinos. It is formed as a result of the dynamical
evolution of a galactic nucleus prior to its collapse into a
massive black hole and formation of high-luminosity AGN. This stage
can be referred to as ``pre-AGN''. A dense central stellar cluster
in the galactic nucleus on the late stage of evolution consists of
compact stars (neutron stars and stellar mass black holes). This
cluster is sunk deep into massive gas envelope produced by
destructive collisions of a primary stellar population. Frequent
collisions of neutron stars in a central stellar cluster are
accompanied by the generation of ultrarelativistic fireballs and
shock waves. These repeating fireballs result in a formation of the
expanding rarefied cavity inside the envelope. The charged
particles are effectively accelerated in the cavity and, due to
$pp$-collisions in the gas envelope, they produce high energy
neutrinos. All high energy particles, except neutrinos, are
absorbed in the thick envelope. Duration of this pre-AGN phase is
$\sim10$~yr, the number of the sources can be $\sim 10$ per
cosmological horizon. High energy neutrino signal can be detected
by underground neutrino telescope with effective area
$S\sim1$~km$^2$.


\end{abstract}

\section{Introduction: Hidden Sources}

High energy (HE) neutrino radiation from astrophysical sources is
accompanied by other types of radiation, most notably by the HE
gamma-radiation. These HE gamma-radiation can be used to put upper
limit on the neutrino flux emitted from a source. For example, if
neutrinos are produced due to interaction of HE protons with low
energy photons in extragalactic space or in the sources transparent
for gamma-radiation, the upper limit on diffuse neutrino flux
$I_{\nu}(E)$ can be derived from e-m cascade radiation. This
radiation is produced due to collisions with photons of microwave
radiation $\gamma_{bb}$, such as $\gamma+\gamma_{bb} \to e^++e^-,
\quad e+\gamma_{bb} \to e'+\gamma'$~etc. These cascade processes
transfer the energy density released in high energy photons
$\omega_{\gamma}$ into energy density of the remnant cascade
photons $\omega_{cas}$. These photons get into the observed energy
range $100$~MeV$-10$~GeV and their energy density is limited by
recent EGRET observations \cite{EGRET} as
$\omega_{cas}\leq2\cdot10^{-6}$~eV/cm$^3$. Introducing the energy
density for neutrinos with individual energies higher than E,
$\omega_{\nu}(>E)$, it is easy to obtain the following chain of
inequalities (reading from left to write)
\begin{equation}
\omega_{cas}>  \omega_{\nu}(>E)
= \frac{4\pi}{c}\int\limits_E^{\infty}EI_{\nu}(E)dE >
\frac{4\pi}{c}E\int\limits_E^{\infty}I_{\nu}(E)dE
=\frac{4\pi}{c}EI_{\nu}(>E).
\label{cas-lim}
\end{equation}
Now the upper limit on the integral HE neutrino flux can be written
down as
\begin{equation}
I_{\nu}(>E) < \frac{4\pi}{c}\frac{\omega_{cas}}{E}=
4.8\cdot 10^3E_{eV}^{-1}\mbox{
cm}^{-2}\mbox{s}^{-1}\mbox{sr}^{-1}.
\label{u-l}
\end{equation}
However, there can be sources, where accompanying electromagnetic
radiation, such as gamma and X-rays, are absorbed. They are called
``hidden sources'' \cite{book}. Several models of hidden sources
were discussed in the literature.

\begin{itemize}

\item {\em Young SN Shell} \cite{BePr} during time
$t_{\nu}\sim10^3-10^4$~s are opaque for all radiation, but
neutrinos.

\item {\em Thorne-Zytkow Star} \cite{ThZy}, the binary with a pulsar
submerged into a red giant, can emit HE neutrinos while all kinds
of e-m radiation are absorbed by the red giant component.

\item {\em Cocooned Massive Black Hole} (MBH) in AGN \cite{ber81} is
an example of AGN as hidden source: e-m radiation is absorbed in a
cocoon around the massive black hole.

\item {\em AGN with Standing Shock} in the vicinity of a MBH
\cite{Ste} can produce large flux of HE neutrinos with relatively
weak X-ray radiation.

\end{itemize}

In this paper we propose a new model of the hidden source which can
operate in a galactic nucleus at pre-AGN phase, i.e. prior to MBH
formation in it. The MBH in AGN is formed through the dynamical
evolution of a central stellar cluster resulting in a secular
contraction of the cluster and its final collapse
\cite{beree78,ree84}. The first stage of this evolution is
accompanied by collisions and destruction of normal stars in the
evolving cluster, when virial velocities of constituent stars become
high enough. The compact stars (neutron stars and black holes)
survive this stage and their population continue to contract, being
surrounded by the massive envelope composed of the gas from destroyed
normal stars. Pre-AGN phase corresponds to a near collapsing
central cluster of compact stars in the galactic nucleus. Repeating
fireballs after continuous collisions of compact stars in this very
dense cluster result in the formation of a rarefied cavity in the
massive gas envelope. Particles accelerated in this cavity interact
with the gas in the envelope and produce HE neutrinos. Accompanying
gamma-radiation can be fully absorbed in the case of thick envelope
(matter depth $X_{env}\sim10^4$~g/cm$^2$). The proposed source is
short-lived (lifetime $t_s\sim10$~years) and very powerful:
neutrino luminosity exceeds the Eddington limit for e-m radiation.

\section{The Model}

We consider in the following the basic features of the formation of
a short-lived extremely powerful hidden source of HE neutrinos in
the process of dynamical evolution of the central stellar cluster
in a typical galactic nucleus.

\subsection{Dynamical Evolution of Galactic Nucleus}
\label{dynamic}

The dynamical evolution of dense central stellar clusters in the
galactic nuclei is accompanied by a secular growth of the
velocity dispersion of constituent stars $v$ or, equivalently, by
the growth of the central gravitational potential. This process is
terminated by the formation of the MBH when the velocity dispersion
of stars grows up to the light speed (see for a review e.g.
\cite{ree84} and references therein). On its way to a MBH
formation the dense galactic nuclei inevitably proceed through the
stellar collision phase of evolution \cite{spi66}--\cite{dok91},
when most normal stars in the cluster are disrupted in
mutual collisions. The necessary condition for the collisional
destruction of normal stars with mass $m_*$ and radius $r_*$ in the
cluster of identical stars with a velocity dispersion $v$ is
\begin{equation}
v>v_p,
\label{disr}
\end{equation}
where
\begin{equation}
v_p=\left(\frac{2Gm_*}{r_*}\right)^{1/2}
\simeq6.2\cdot10^2\left(\frac{m_*}{\rm M_{\odot}}\right)^{1/2}
\left(\frac{r_*}{R_{\odot}}\right)^{-1/2} \mbox{ km s}^{-1}.
\label{vpar}
\end{equation}
is an escape (parabolic) velocity from the surface of a constituent
normal star. The kinetic energy of colliding star is greater in
general than its gravitational bound energy under the inequality
(\ref{disr}). If $v>v_p$, the normal stars are eventually disrupted
in mutual collisions or in collisions with the extremely compact
stellar remnants, i.e. with neutron stars (NSs) or stellar mass
black holes. Only these compact stellar remnants will survive
through the stellar-destruction phase of evolution ($v=v_p$) and
form the self-gravitating core. We shall refer for simplicity to
this core as to the NS cluster. Meanwhile the remnants of disrupted
normal stars form a gravitationally bound massive gas envelope in
which the NS cluster is submerged. The virial radius of this
envelope is
\begin{equation}
R_{env}=\frac{GM_{env}}{2v_p^2}
=\frac{1}{4}\frac{M_{env}}{m_*}r_* \simeq0.56M_8
\left(\frac{m_*}{\rm M_{\odot}}\right)^{-1}
\left(\frac{r_*}{R_{\odot}}\right) \mbox{ pc},
\label{Renv}
\end{equation}
where $M_{env}=10^8M_8{\rm M_{\odot}}$ is a corresponding mass of
the envelope. The gas from disrupted normal stars composes the
major part of the progenitor central stellar cluster in the
galactic nucleus. So the natural range for the total mass of the
envelope is the same as the typical range for the mass of a central
stellar cluster in the galactic nucleus, $M_{env}=10^7-10^8{\rm
M_{\odot}}$. The envelope radius $R_{env}$ is given by the virial
radius of a central cluster in the galactic nucleus at the moment
of evolution corresponding to normal stars destructions, i.~e,
$v=v_p$. The mean number density of gas in the envelope is
\begin{equation}
n_{env}=\frac{3}{4\pi}\frac{1}{R_{env}^3}\frac{M_{env}}{m_p}
\simeq5.4\cdot10^{9}M_8^{-2}
\left(\frac{m_*}{\rm M_{\odot}}\right)^{3}
\left(\frac{r_*}{R_{\odot}}\right)^{-3} \mbox{ cm}^{-3},
\label{nenv}
\end{equation}
where $m_p$ is a proton mass. A column density of the envelope is
\begin{equation}
X_{env}=m_p n_{env} R_{env}
\simeq1.6\cdot10^4M_8^{-1}
\left(\frac{m_*}{\rm M_{\odot}}\right)^{2}
\left(\frac{r_*}{R_{\odot}}\right)^{-2} \mbox{ g cm}^{-2}.
\label{depth}
\end{equation}
Such an envelope completely absorbs electromagnetic radiation and
HE particles outgoing from the interior, except neutrinos and
gravitational waves. A column density becomes less for more massive
envelopes.

We assume that energy release due to star collisions supports the
gas in the cluster in (quasi-)dynamical equilibrium. It implies the
equilibrium temperature $T_{eq}$ of the gas,
\begin{equation}
T_{eq} \sim \frac{m_p}{6k}\frac{GM_{env}}{R_{env}} \sim
1.6\cdot 10^7\mbox{ K}.
\label{Teq}
\end{equation}
The thermal velocity of gas particles at the equilibrium
temperature is of order of an escape velocity from the surface of a
normal star.

\noindent
If $T \gg T_{eq}$ the gas outflows from the cluster with the sound speed
\begin{equation}
v_s=\left( 2\gamma \frac{kT}{m_p} \right)^{1/2}
\approx 600\left( \frac{T}{T_{eq}}\right)^{1/2}\mbox{ km/s},
\label{vsound}
\end{equation}
where $\gamma$ is adiabatic index ($\gamma=5/3$ for hydrogen).\\
\noindent
If $T \ll T_{eq}$ gas collapses to the core.

\subsection{Dense Cluster of Stellar Remnants}

As was discussed above, the dense NS cluster survives inside the
massive envelope of the post-stellar-destruction galactic nucleus.
The total mass of this cluster is $\sim1-10$~\% of the total mass
of a progenitor galactic nucleus \cite{spi66}--\cite{spi71} and so
of the massive envelope, i.~e. $M\sim0.01 - 0.1M_{env}$. We will
use the term `evolved galactic nucleus' for this cluster of NSs
assuming that (i) $v>v_p$ and (ii) the (two-body) relaxation time
in the cluster is much less than the age of the host galaxy. Under
the last condition the cluster has enough time for an essential
dynamical evolution. For example the relaxation time of stars
inside a central parsec of the Milky Way galaxy is
$t_r\sim10^7-10^8$~years. A further dynamical evolution of the
evolved cluster is terminated by the dynamical collapse to a MBH.

We consider in the following an evolved central cluster of NSs with
identical masses $m=1.4{\rm M_{\odot}}$. This evolved cluster of
NSs is sunk deep into the massive gas envelope remaining after the
previous evolution epoch of a typical normal galactic nucleus. Let
$N=M/m=10^6N_6$ is a total number of NSs stars in the cluster. The
virial radius of this cluster is:
\begin{equation}
R=\frac{GNm}{2v^2}=\frac{1}{4}\left(\frac{c}{v}\right)^2 N r_g
\simeq1.0\cdot10^{13} N_6 (v/0.1c)^{-2} \mbox{ cm},
\label{Radius}
\end{equation}
where $r_g=2Gm/c^2$ is a gravitational radius of NS. For
$N\sim10^6$ and $v\sim0.1c$ one has nearly collapsing cluster with
the virial size of $\sim1$~AU. The characteristic times are (i) the
dynamical time
$t_{dyn}=R/v=(1/4)N(c/v)^{3}r_g/c\simeq0.95N_6(v/0.1c)^{-3}$~hour
and (ii) the evolution (two-body relaxation) time of the NS cluster
$t_{rel}\simeq0.1 (N/\ln
N)t_{dyn}\simeq19N_6^2(v/0.1c)^{-3}$~years. In general $t_{rel}\gg
t_{dyn}$, if $N\gg 1$. This evolution time determines the duration
of an active phase for the considered below hidden source, as
$t_s\sim t_{rel}\sim10$~years.

\subsection{Fireballs in Cluster}

The most important feature of our model is a secular growing rate
of accidental NS collisions in the evolving cluster, accompanied by
large energy release. The corresponding rate of NS collisions in
the cluster (with the gravitational radiation losses taken into
account) is \cite{qui87}--\cite{dok98}
\begin{eqnarray}
\dot N_c & = & 9\sqrt2\left(\frac{v}{c}\right)^{17/7}\frac{c}{R}=
36\sqrt2\left(\frac{v}{c}\right)^{31/7}\frac{1}{N}\frac{c}{r_g}
\nonumber \\
& \simeq & 4.4\cdot10^{3}N_6^{-1}(v/0.1c)^{31/7} \mbox{
yr}^{-1}.
\label{Nrate}
\end{eqnarray}
The time between two successive NS collisions is
\begin{eqnarray}
t_c & = & \dot N^{-1}_c
=\frac{1}{9\sqrt2}\left(\frac{c}{v}\right)^{10/7}t_{dyn}
=\frac{1}{36\sqrt2}\left(\frac{c}{v}\right)^{31/7}
N\frac{r_g}{c}\nonumber \\ & \simeq &
7.3\cdot10^{3}N_6(v/0.1c)^{-31/7} \mbox{ s}.
\label{tcoll}
\end{eqnarray}
Note that a number of NS collisions, given by $\dot{N_c}t_s$,
comprises only a small fraction, about $1\%$, of a total number of
NSs in the cluster to the time of the onset of dynamical collapse
of the whole cluster into a MBH.

Merging of two NSs in collision is similar to the process of a
tight binary merging: the NSs are approaching to each other by
spiralling down due to the gravitational waves radiation and then
coalesce by producing an ultrarelativistic photon-lepton fireball
\cite{cav78}--\cite{mes92}, which we assume to be spherically
symmetric.

The energy of one fireball is $E_0=E_{52}10^{52}$~ergs, and the
total energy release in the form of fireballs during lifetime of
the hidden source $t_s\sim10$~yr is
\begin{equation}
E_{tot} \sim \dot{N_c} E_0 t_s \sim 4\cdot 10^{56}\mbox{ ergs},
\label{Etot}
\end{equation}
where $\dot{N_c}$ is the NS collision rate.

The physics of fireballs is extensively elaborated especially in
recent years for the modeling of cosmological gamma-ray bursts
(GRBs) (for review see e.~g. \cite{pir96} and references therein).
The newborn fireball expands with relativistic velocity,
corresponding to the Lorentz factor $\Gamma_f\gg 1$. The relevant
parameter of a fireball is the total baryonic mass
\begin{equation}
M_0= E_0/\eta c^2\simeq5.6\cdot10^{-6}E_{52}\eta_3^{-1}M_{\odot},
\label{M0}
\end{equation}
where baryon-loading mass parameter $\eta=10^3\eta_3$. The maximum
possible Lorentz factor of expanding fireball is $\Gamma_f=\eta+1$
during the matter-dominated phase of fireball expansion
\cite{cav78,she90}. During the initial phase of expansion, starting
from the radius of the `inner engine' $R_0\sim10^6 - 10^7$~cm, the
fireball Lorentz factor increases as $\Gamma\propto r$, until it is
saturated at the maximum value $\Gamma_f=\eta\gg1$ at the radius
$R_{\eta}=R_0\eta $ (see e.~g. \cite{pir96}). Internal shocks will
take place around $R_{sh}=R_0\eta^2$, if the fireball is
inhomogeneous and the velocity is not a monotonic function of
radius, e.g. due to the considerable emission fluctuations of the
inner engine \cite{pac94}--\cite{kob97}. The fireball expands with
the constant Lorentz factor $\Gamma=\eta$ at $R>R_{\eta}$ until it
sweeps up the mass $M_0/\eta$ of ambient gas and looses half of its
initial momentum. At this moment ($R=R_{\gamma}$) the deceleration
stage starts \cite{mes92}--\cite{mes93}.

Interaction of the fireball with an ambient gas determines the
length of its relativistic expansion. In our case the fireball
propagates through the massive envelope with a mean gas number
density $n_{env}=\rho_{env}/m_p=n_9 10^9$~cm$^{-3}$ as it follows
from Eq.~(\ref{nenv}). Fireball expands with $\Gamma\gg1$ up to the
distance determined by the Sedov length
\begin{equation}
l_S=\left(\frac{3}{4\pi}\frac{E_0}{\rho_{env}c^2}\right)^{1/3}
\simeq1.2\cdot10^{15}n_9^{-1/3}E_{52}^{1/3} \mbox{ cm}.
\label{Sedov}
\end{equation}
Fireball becomes mildly relativistic at radius $r=l_S$ due to
sweeping up the gas from the envelope with the mass $M_0 \eta$. The
radius $r=l_S$ is the end point of the ultrarelativistic fireball
expansion phase. Far beyond the Sedov length radius ($r\gg l_S$)
there is the non-relativistic Newtonian shock driven by the
decelerated fireball. Its radius $R(t)$ obeys the Sedov--Taylor
self-similar solution \cite{lan87}, with
$R(t)=(E_0t^2/\rho_{env})^{1/5}$. The corresponding shock expansion
velocity is $u=(2/5)[l_S/R(t)]^{3/2}c\ll c$.

\subsection{Cavity and Shocks}

We show here that relativistic fireballs from a dense central
cluster of NSs produce the dynamically supported rarefied cavity
deep inside the massive gaseous envelope.

The first fireball sweeps out the gas from the envelope producing
the cavity with a radius $l_S$. This cavity expands due to the next
fireballs, which propagate first in a rarefied cavity and then hit
the boundary pushing it further. 

Each fireball hitting the dense
envelope is preceded by a shock. Propagating through the envelope,
the shock sweeps up the gas ahead of it and gradually decelerates.
The swept out gas forms a thin shell with a density profile given
by Sedov self-similar solution. The next fireball hits this thin
shell when it is decelerated down to the non-relativistic velocity.
Moving in the envelope, the shell accumulates more gas,
retaining the same density profile, and then it is hit by the next
fireball again. After a number of collisions the shell becomes massive,
and the successive hitting fireballs do not change its velocity
appreciably. In this regime one can consider the propagation of
massive non-relativistic thin shell with a shock (density
discontinuity) ahead of it. The shock speed $v_{sh}$ is connected
with a velocity $v_g$ of gas behind it as $v_{sh}=(\gamma+1)v_g/2$,
where $\gamma$ is an adiabatic index. The density perturbation in
the envelope propagates as a shock until $v_{sh}$ remains higher
than sound speed $v_s$. In the considered case, the shock
dissipates in the middle of the envelope. For the Sedov solution the
shock velocity changes with distance $r$ as
\begin{equation}
v_{sh}(r)=
\frac{2}{5}\alpha_S^{-1}\left(\frac{E_{sh}}{\rho_{env}}\right)^{1/2}
r^{-3/2},
\label{v-sh}
\end{equation}
where $\alpha_S$ is the constant of self-similar Sedov solution;
when radiative pressure dominates $\alpha_S=0.894$. The other
quantities in Eq.~(\ref{v-sh}) are $E_{sh}= (1/2)E_{tot} = 2\cdot
10^{56}$~erg/s is the energy of the shock, which includes kinetic
and thermal energy (the half of a total energy is transformed to
particles accelerated in the cavity), and $\rho_{env}$ is a density
of the envelope given by Eq.~(\ref{nenv}). From Eqs.~(\ref{v-sh})
and (\ref{vsound}) it follows that $v_{sh}>v_s$ holds at distance
$r<1\cdot10^{18}\mbox{ cm}\sim0.6R_{env}$, i.~e. that shock does
not reach the outer surface of the envelope. In fact the latter
conclusion follows already from energy conservation. The
gravitational energy of the envelope $V_0$ is
\begin{equation}
V_0=\kappa \frac{GM_{env}^2}{R_{env}}>9.2\cdot 10^{56}\mbox{ ergs},
\label{V_0}
\end{equation}
where $\kappa$ depends on a radial profile of the gas density in
the envelope $\rho(r)$, and changes from 3/5 to 1. This energy is
higher than a total injected energy $E_{tot}\sim 4\cdot
10^{56}$~ergs, and thus the system remains gravitationally bound.
When a shock reaches the boundary of the envelope, the gas
distribution changes there. It has the form of a thin shell with
gravitational energy $V_e=GM_{env}^2/2R_{env}$. To provide the exit
of the shock to the surface of the envelope,
a total energy release must satisfy the relation $E_{tot}> V_0-V_e$, 
where the minimum value of $V_0-V_e$ is 
$1.5\cdot 10^{56}$~ergs for $\kappa=3/5$. Actually $E_{tot}$ must 
be higher because (i) part of the
injected energy goes to heat, (ii) more realistic value of $\kappa
=1$, and (iii) the shell has a non-zero velocity, when the shock
disappears, with a gravitational braking taken into account. We
conclude thus that for $E_{tot}\sim 4\cdot 10^{56}$~ergs, shock
dissipates inside the envelope.

The cavity radius grows with time. For the stage, when a shell
moves non-relativistically, the cavity radius calculated as a
distance to the shell in the Sedov solution is
\begin{equation}
R_{cav}(t)=\left(\frac{E_{sh}}{\alpha_S \rho_{env}}\right )^{1/5}t^{2/5}.
\label{R-cav}
\end{equation}
At the end of a phase of the hidden source activity,
$t_s\sim10$~yr, the radius of the cavity reaches
$\sim3\cdot10^{17}$~cm, remaining thus much less than $R_{env}$.

The cavity is filled by direct and reverse relativistic shocks from
fireballs. Reverse shocks are produced by decelerated fireballs,
most notably when they hit the boundary of the cavity. 
The expanding fireballs inside the cavity have a
shape of thin shells \cite{MeLaRe} and are separated by distance
\begin{equation}
R_c= ct_c\simeq2.2\cdot10^{14}N_6(v/0.1c)^{-31/7}\mbox{ cm}.
\label{Rc}
\end{equation}
The gas between two fireballs is swept up by the preceding one. A
total number of fireballs existing in the cavity simultaneously is
$N_f \sim R_{cav}/R_c\gg1$, and this number grows with time as
$t^{2/5}$.

Shocks generated by repeating fireballs are ultrarelativistic
inside the cavity and mildly relativistic in the envelope near the
inner boundary. Collisions of multiple shocks in the cavity, as
well as inside the fireballs, produce strongly turbulized medium
favorable for generation of magnetic fields and particle
acceleration.

\section{High Energy Particles in Cavity}

There are three regions where acceleration of particles take place.

(i) NS cluster, where fireballs collide, producing the turbulent
medium with large magnetic field. This region has a small size of
order of virial radius of the cluster $R \sim 10^{13}$~cm, and we
neglect its contribution to production of accelerated particles.

(ii) The region at the boundary between the cavity and the
envelope. During the active period of a hidden source, $t_s \sim
10$~yr, the fireballs hit this region, heating and turbulizing it.
The large magnetic equipartition field is created here. This
boundary region has density $\rho \sim \rho_{env}$, the radius $R
\sim R_{cav}$ and the width $\Delta < 0.1 R_{cav}$.

(iii) Most of the cavity volume is occupied by fireballs, separated
by distance $R_c$. Due to collisions of internal shocks, the gas in
a fireball is turbulized and equipartition magnetic field is
generated \cite{pir96,wax}.

In all three cases the Fermi II acceleration mechanism operates.
For all three sites we assume existence of equipartition magnetic
field, induced by turbulence and dynamo mechanism:
\begin{equation}
\frac{H^2}{8\pi} \sim \frac{\rho u_t^2}{2},
\label{equip}
\end{equation}
where $\rho$ and $u_t$ are the gas density and velocity of
turbulent motions in the gas, respectively. Since turbulence is
caused by shocks, the shock spectrum of turbulence $F_k \sim
k^{-2}$ is valid, where $k$ is a wave number. Assuming
equipartition magnetic field on each scale $l \sim 1/k, \quad
H_l^2\sim kF_k$, one obtains the distribution of magnetic fields
over the scales as
\begin{equation}
H_l/H_0 = (l/l_0)^{1/2},
\end{equation}
where $l_0$ is a maximum scale with the coherent field $H_0$ there.

The maximum energy of accelerated particles is given by
\begin{equation}
E_{max} \sim eH_0l_0
\label{Emax}
\end{equation}
with an acceleration time
\begin{equation}
t_{acc} \sim \frac{l_0}{c} \left( \frac{c}{v}\right)^2.
\label{tac}
\end{equation}
For the turbulent shell at the boundary between the cavity and
envelope, assuming mildly relativistic turbulence $u_t \sim c$ and
$\rho \sim \rho_{env}$, we obtain $H_{eq}=4\cdot10^3$~G. The
maximum acceleration energy is $E_{max}= 2\cdot 10^{21}$~eV, if the
coherent length of magnetic field $l_0$ is given by the Sedov
length $l_{S}$, and the acceleration time is $t_{acc}= 4\cdot 10^4
E_{52}^{1/3}n_9^{-1/3}$~s. The typical time of energy losses,
determined by $pp$-collisions, is much longer than $t_{acc}$, and
does not prevent acceleration to $E_{max}$ given above:
\begin{equation}
t_{pp}=\left(\frac{1}{E}\frac{dE}{dt}\right)^{-1}
=\frac{1}{f_p\sigma_{pp}n_{env} c}
\simeq2\cdot10^6 n_9^{-1} \mbox{ s},
\label{tpp}
\end{equation}
where $f_p \approx 0.5$ is the fraction of energy lost by HE proton
in one collision, $\sigma_{pp}$ is a cross-section of
$pp$-interaction, and $n_{env}$ is the gas number density in the
boundary turbulent shell.

The turbulence in the fireball is produced by collisions of
internal shells, and a natural scale for coherence length $l_0$ is
a width of the internal shell in the local frame $\delta'$. The
maximum energy in the laboratory frame $E_{max}\sim
eH'_{eq}\delta$, where $\delta$ is the corresponding width in the
laboratory frame. Since $H'\propto 1/R$ and $\delta \propto R$, the
maximum energy does not change with time, and can be estimated as
in Ref.~\cite{wax}, $E_{max}\sim 3\cdot 10^{20}$~eV. Note that in
our case a fireball propagates in the very low-density medium.

A gas left in the cavity by preceding fireball, as well as high
energy particles escaping from it, are accelerated by the next
fireball by a factor $\Gamma_f^2$ at each collision. This
$\Gamma^2$-mechanism of acceleration works only in pre-hydrodynamic
regime of fireball expansion, after reaching the hydrodynamic
stage, $\Gamma^2$-mechanism ceases \cite{BBH}.

We conclude, thus, that both efficiency and maximum acceleration
energy are very high. We assume that a fireball transfers half of
its energy to accelerated particles.

\section{Neutrino Production and Detection}

Particles accelerated in the cavity interact with the gas in the
envelope producing high energy neutrino flux. We assume that about
half of the total power of the source $L_{tot}$ is converted into
energy of accelerated particles $L_p \sim 7\cdot 10^{47}$~erg/s. As
estimated in Section 2.1, the column density of the envelope varies
from $X_{env}\sim10^2$~g/cm$^2$ (for very heavy envelope) up to
$X_{env}\sim10^4$~g/cm$^2$ (for the envelope with mass $M \sim
10^8M_{\odot}$. Taking into account the magnetic field, one
concludes that accelerated protons loose in the envelope a large
fraction of their energy. The charged pions, produced in
$pp$-collisions, with Lorentz factors up to
$\Gamma_c\sim1/(\sigma_{\pi N}n_{env}c\tau_{\pi})
\sim4\cdot10^{13}n_9^{-1}$ freely decay in the envelope (here
$\sigma_{\pi N}\sim3\cdot10^{-26}$~cm$^2$ is $\pi N$-cross-section,
$\tau_{\pi}$ is the lifetime of charged pion, and $n_{env}=10^9
n_9$~cm$^{-3}$ is the number density of gas in the envelope). We
assume $E^{-2}$ spectrum of accelerated protons
\begin{equation}
Q_p(E)= \frac{L_p}{\zeta E^2},
\end{equation}
where $\zeta=\ln (E_{max}/E_{min})\sim20-30$. About half of its
energy protons transfer to high energy neutrinos through decays of
pions, $L_{\nu} \sim (2/3)(3/4)L_p$, and thus the production rate
of $\nu_{\mu}+\bar{\nu}_{\mu}$ neutrinos is
\begin{equation}
Q_{\nu_{\mu}+\bar{\nu}_{\mu}}(>E) = \frac{L_p}{4\zeta E^2}.
\label{Qnu}
\end{equation}
Crossing the Earth, these neutrinos create deep underground the
equilibrium flux of muons, which can be calculated as \cite{nu90}:
\begin{equation}
F_{\mu}(>E)=\frac{\sigma_0
N_A}{b_{\mu}}Y_{\mu}(E_{\mu})\frac{L_p}{4\xi E_{\mu}}
\frac{1}{4\pi r^2} , \label{Fmu}
\end{equation}
where the normalization cross-section
$\sigma_0=1\cdot10^{-34}$~cm$^2$, $N_A=6\cdot10^{23}$ is the
Avogadro number, $b_{\mu}=4\cdot10^{-6}$~cm$^2$/g is the rate of
muon energy losses, $Y_{\mu}(E)$ is the integral muon moment of
$\nu_{\mu}N$ interaction (see e.~g. \cite{book,nu90}). The most
effective energy of muon detection is $E_{\mu}\geq1$~TeV
\cite{nu90}. The rate of muon events in the underground detector
with effective area $S$ at distance $r$ from the source is given by
\begin{equation}
\dot N(\nu_{\mu})=F_{\mu}S\simeq70\left(\frac{L_p}{10^{48}\mbox{
erg s}^{-1}}\right)
\left(\frac{S}{1\mbox{ km}^{2}}\right)
\left(\frac{r}{10^3\mbox{ Mpc}}\right)^{-2}
\quad\mbox{yr}^{-1}. \label{Number}
\end{equation}
Thus, we expect about 10 muons per year from the source at distance
$10^3$~Mpc.

\section{Accompanying Radiation}

We shall consider below HE gamma-ray radiation produced by
accelerated particles and thermalized infrared radiation from the
envelope. As far as HE gamma-ray radiation is concerned, there will
be considered two cases: (i) thin envelope with
$X_{env}\sim10^2$~g/cm$^2$ and (ii) thick envelope with
$X_{env}\sim10^4$~g/cm$^2$. In the latter case HE gamma-ray
radiation is absorbed.

\subsection{Gamma-Ray Radiation}

Apart from high energy neutrinos, the discussed source can emit HE
gamma-radiation through $\pi^0\to2\gamma$ decays and synchrotron
radiation of the electrons. In case of the thick envelope with
$X_{env}\sim10^4$~g/cm$^2$ most of HE photons are absorbed in the
envelope (characteristic length of absorption is the radiation
length $X_{rad}\approx60$~g/cm$^2$). In the case of the thin
envelope, $X_{env}\sim100$~g/cm$^2$, HE gamma-radiation emerges
from the source. Production rate of the synchrotron photons can be
readily calculated as
\begin{equation}
dQ_{syn}=\frac{dE_e}{E_{\gamma}}Q_e(>E_e),
\end{equation}
where $E_e$ and $E_{\gamma}$ are the energies of electron and of
emitted photon, respectively. Using $Q_e(>E_e)=L_e/(\eta E_e)$ and
$E_{\gamma}=k_{syn}(H)E_e^2$, where $k_{syn}$ is the coefficient of
the synchrotron production, one obtains
\begin{equation}
Q_{syn}(E_{\gamma})= \frac{1}{12} \frac{L_p}{\eta E_{\gamma}^2}.
\label{Qsyn}
\end{equation}
Note, that the production rate given by Eq.~(\ref{Qsyn}) does not
depend on magnetic field. Adding the contribution from $\pi^0 \to
2\gamma$ decays, one obtains
\begin{equation}
Q_{\gamma}(E_{\gamma})= \frac{5}{12} \frac{L_p}{\eta E_{\gamma}^2},
\label{Qgamma}
\end{equation}
and the flux at $E_{\gamma}\geq1$~GeV at the distance to the source
$r=1\cdot10^3$~Mpc is
\begin{eqnarray}
F_{\gamma}(>E_{\gamma}) & = &
\frac{5}{12} \left(\frac{1}{4\pi
r^2}\right)\frac{L_p}{\xi E_{\gamma}} \nonumber\\
& \simeq &
2.2\cdot 10^{-8}\left(\frac{L_p}{10^{47}\mbox{ erg s}^{-1}}\right)
\left(\frac{r}{10^3\mbox{ Mpc}}\right)^{-2}
\mbox{ cm}^{-2} \mbox{ s}^{-1},
\end{eqnarray}
i.~e. the source is detectable by EGRET.

\subsection{Infrared and Optical Radiation}

Hitting the envelope, fireballs dissipate part of its kinetic
energy in the envelope in the form of low-energy e-m radiation.
This radiation is thermalized in the optically thick envelope and
then re-emitted in the form of black-body radiation from the
surface of the envelope. It appears much more later than HE
neutrino and gamma-radiation. The thermalized radiation diffuses
through the envelope with a diffusion coefficient $D \sim c
l_{dif}$, where a diffusion length is $l_{dif}=1/(\sigma_T
n_{env})$ and $\sigma_T$ is the Thompson cross-section. A mean time
of the radiation diffusion through the envelope of radius $R_{env}$
is
\begin{equation}
t_d \sim \frac{R_{env}^2}{D} \sim 1\cdot 10^4\mbox{ yr},
\label{tdiff}
\end{equation}
independently of the envelope mass. This diffusion time is to be
compared with a duration of an active phase $t_s \sim 10$~years
and with a time of flight $R_{env}/c \sim2$~years.

Since the produced burst is very short, $t_s \sim 10$~yr, the
arrival times of photons to the surface of the envelope have a
distribution with a dispersion $\sigma \sim t_d$. An average
surface black body luminosity is then
\begin{equation}
\bar{L}_{bb}\sim E_{tot}/t_d \sim 1\cdot10^{45}\mbox{ erg/s},
\label{Lbb}
\end{equation}
with a peak luminosity being somewhat higher. The temperature of
this radiation corresponds to IR range
\begin{equation}
T_{bb}=
\left(\frac{\bar{L}_{bb}}{4\pi R^2_{env}\sigma_{SB}}\right)^{1/4}
\simeq 8.4\cdot10^2\mbox{ K}.
\end{equation}
Thus, in $\sim 10^4$ years after neutrino burst, the hidden source
will be seen in the sky as a luminous IR source. Since we consider
a model of production of high-luminosity AGN, the object is
typically expected to be at high redshift $z$. Its visible
magnitude is
\begin{equation}
m=-2.5\log \left(\frac{L_{bb}H_0^2}{16\pi(z+1-\sqrt{1+z})^2f_0}\right)
\label{magn}
\end{equation}
where $H_0=100h$~km/s~Mpc is the Hubble constant and the flux
$f_0=2.48\cdot 10^{-5}$~erg/cm$^2$s. For $L_{bb}=1\cdot
10^{45}$~erg/s, $z=3$ and $h=0.6$, the visible magnitude of the IR
source is $m=22.7$. Such a faint source is not easy identify in
e.g. IRAS catalogue as a powerful source, because for redshift
determination it is necessary to detect the optical lines from the
host galaxy, which are very weak at assumed redshift $z=3$. The
non-thermal optical radiation can be also produced due to HE
proton-induced pion decays in the outer part of the envelope, but
its luminosity is very small. Most probably such source will be
classified as one of the numerous non-identified weak IR source.

\subsection{Duration of activity and the number of sources}

As was indicated in Section 2.2 the duration of the active phase
$t_s$ is determined by relaxation time of the NS cluster: $t_s\sim
t_{rel}\sim10-20$~yr. This stage appears only once during the
lifetime of a galaxy, prior to the MBH formation. If to assume that
a galactic nucleus turns after it into AGN, the total number of
hidden sources in the Universe can be estimated as
\begin{equation}
N_{HS} \sim \frac{4}{3}\pi (3ct_0)^3 n_{AGN} t_s/t_{AGN},
\label{totnum}
\end{equation}
where $\frac{4}{3}\pi(3ct_0)^3$ is the cosmological volume inside
the horizon $ct_0$, $n_{AGN}$ is the number density of AGNs and
$t_{AGN}$ is the AGN lifetime. The value $t_s/t_{AGN}$ gives a
probability for AGN to be observed at the stage of the hidden
source, if to include this short stage ($t_s \sim 10$~yr) in the
much longer ($t_{AGN}$) AGN stage and consider (for the aim of
estimate) the hidden source stage as the accidental one in the AGN
history. The estimates for $n_{AGN}$ and $t_{AGN}$ taken for
different populations of AGNs result in $N_{HS} \sim 10 - 100$.

\section{Conclusions}

Dynamical evolution of the central stellar cluster in the galactic
nucleus results in the stellar destruction of the constituent
normal stars and in the production of massive gas envelope. The
surviving subsystem of NSs submerges deep into this envelope. The
fast repeating fireballs caused by NS collisions in the central
stellar cluster produce the rarefied cavity inside the massive
envelope. Colliding shocks generate the turbulence inside the
fireballs and in the cavity, and particles are accelerated by Fermi
II mechanism. These particles are then re-accelerated by
$\Gamma^2$-mechanism in collisions with relativistic shocks and
fireballs.

All high energy particles, except neutrinos, can be completely
absorbed in the thick envelope. In this case the considered source
is an example of a powerful hidden source of HE neutrinos.

Prediction of high energy gamma-ray flux depends on the thickness
of envelope. In case of the thick envelope,
$X_{rad}\sim10^4$~g/cm$^2$, HE gamma-radiation is absorbed. When an
envelope is thin, $X_{rad}\sim10^2$~g/cm$^2$, gamma-ray radiation
from $\pi^0\to2\gamma$ decays and from synchrotron radiation of the
secondary electrons can be observed by EGRET and marginally by
Whipple detector at $E_{\gamma}\geq1$~TeV.

In all cases the thickness of the envelope is much larger than the
Thompson thickness ($x_T\sim3$~g/cm$^{2}$), and this condition
provides the absorption and X-rays and low energy gamma-rays.

A hidden source is to be seen as a bright IR source but, due to
slow diffusion through envelope, this radiation appears
$\sim10^4$~years after the phase of neutrino activity. During the
period of neutrino activity the IR luminosity is the same as before
it. A considered source is a precursor of most powerful AGN, and
therefore most of these sources are expected to be at the same
redshifts as AGNs. The luminosity $L_{IR} \sim 10^{45} -
10^{46}$~erg/s is not unusual for powerful IR sources from IRAS
catalogue. The maximum observed luminosity exceeds
$1\cdot10^{48}$~erg/s \cite{Row}, and there are many sources with
luminosity $10^{45}-10^{46}$~erg/s \cite{SaMi}. Moreover, for
most of the hidden sources the distance cannot be determined, and
thus they fall into category of faint non-identified IR sources.

Later these hidden sources turn into usual powerful AGNs, and
thus the number of hidden sources is restricted by the total
number of these AGNs.

In our model the shock is fully absorbed in the envelope. Since
the total energy release $E_{tot}$ is less than gravitational energy
of the envelope $E_{grav} \sim GM_{env}^2/R_{env}$, the system remains
gravitationally bound, and in the end the envelope will collapse
into black hole or accretion disc.

The expected duration of neutrino activity for a hidden source is
$\sim10$~yr, and the total number of hidden sources in the horizon
volume ranges from a few up to $\sim100$, within uncertainties of
the estimates.

Underground neutrino detector with an effective area
$S\sim1$~km$^2$ will observe $\sim10$ muons per year with energies
$E_{\mu}\geq1$~TeV from this hidden source.

{\bf Acknowledgments:} We are grateful to Bohdan Hnatyk for useful
discussions. This work was supported in part by the INTAS through
grant No. 99-1065. One of the authors (VID) is grateful to the
staff of Laboratori Nazionali del Gran Sasso for hospitality during
his visit.


\begin{thebibliography}{99}
\bibitem{EGRET} P. Sreekumar et al., (EGRET collaboration),
        Astrophys. J. 494 (1998) 523.
\bibitem{book} V.S. Berezinsky, S.V. Bulanov, V.A. Dogiel, V.L. Ginzburg,
        V.S.~Ptu\-skin, Astrophysics of Cosmic Rays, (North-Holland,
        Amsterdam, 1990).
\bibitem{BePr} V.S. Berezinsky, O.F. Prilutsky,  Astron. \& Astrophys.
        66 (1987) 325.
\bibitem{ThZy} K.S. Thorne, A.N. Zytkow, Astrophys. J. 212 (1977) 832.
\bibitem{ber81} V.S. Berezinsky, V.L. Ginzburg,  Mon. Not. R. Astron. Soc.
        194 (1981) 3.
\bibitem{Ste} F.W. Stecker, C.Done, M.H. Salamon, P. Sommers,
        Phys. Rev. Lett. 66 (1991) 2697.
\bibitem{beree78} M.C. Begelman, M.J. Rees Mon. Not. R. Astron. Soc.
185 (1978) 847.
\bibitem{ree84} M.J. Rees,  Ann. Rev. Astron. \& Astrophys. 22 (1984) 471.
\bibitem{spi66} L. Spitzer, W.C. Saslaw, Astrophys. J. 143 (1966) 400.
\bibitem{col67} S.A. Colgate, Astrophys. J. 150 (1967) 163.
\bibitem{san70} R.H. Sanders, Astrophys. J. 162 (1970) 791.
\bibitem{spi71} L. Spitzer,  Galactic Nuclei, D.~Q'Connel, ed.
        (North Holland, Amsterdam, 1971), p.~443.
\bibitem{dok91} V.I. Dokuchaev, Mon. Not. R. Astron. Soc. 251 (1991) 564.
\bibitem{qui87} C.D. Quinlan, S.L. Shapiro,  Astrophys. J. 321 (1987) 199.
\bibitem{qui90} C.D. Quinlan, S.L. Shapiro,  Astrophys. J. 356 (1990) 483.
\bibitem{dok98} V.I. Dokuchaev, Yu.N. Eroshenko, L.M. Ozernoy,
        Astrophys. J. 502 (1998) 192.
\bibitem{cav78} G. Cavallo, M.J. Rees, Mon. Not. R. Astron. Soc.
        183 (1978) 359.
\bibitem{goo86} J. Goodman, Astrophys. J. 308 (1986) L47.
\bibitem{pac86} B. Paczy\'nski, Astrophys. J. 308 (1986) L51.
\bibitem{she90} A. Shemi, T. Piran, Astrophys. J. 365 (1990) L55.
\bibitem{mes92} P. M\'esz\'aros, M.J. Rees, Mon. Not. R. Astron. Soc.
        248 (1992) 41P.
\bibitem{pir96} T. Piran, Physics Report, 314 (1999) 575.

\bibitem{pac94} B. Paczy\'nski, G. Hu, Astrophys. J. 427 (1994) 708.
\bibitem{mes94} P. M\'esz\'aros, M.J. Rees, Astrophys. J. 430 (1994) L93.
\bibitem{kob97} S. Kobayashi, T. Piran, R. Sari, Astrophys. J. 490 (1997)
92.
\bibitem{bla76} R.D. Blandford, C.F. McKee, Phys. Fluids, 19 (1976) 1130.
\bibitem{mes93} P. M\'esz\'aros, M.J. Rees, Astrophys. J. 405 (1993) 278.
\bibitem{lan87} L.D. Landau, E.M. Lifshitz, Fluid Mechanics 2nd ed.
        (Pergamon Press, 1987), Chap.~X.
\bibitem{MeLaRe} P.Meszaros, P.Laguna, M.Rees, Astrophys. J. 415 (1993) 181.
\bibitem{wax} E. Waxman, Phys. Rev. Lett. 75 (1995) 386.
\bibitem{BBH} V. Berezinsky, P. Blasi, B. Hnatyk, in preparation.
\bibitem{nu90} V. Berezinsky, Nucl. Phys. B (Proc. Suppl.). 19 (1990) 375.
\bibitem{Row} M. Rowan-Robinson, T. Broadhurst, A. Lawrence, R.G. McMahon
et al, Nature, 351 (1991) 719.
\bibitem{SaMi} D.B. Sanders, I.F. Mirabel, Ann.~Rev.~Astron.~Astroph.
34 (1996) 749.
\end{thebibliography}
\end{document}